\documentclass[twocolumn, trackchanges]{aastex631}
\usepackage{amsmath}

\begin{document}

\title{GRB Progenitor Classification from Gamma-Ray Burst Prompt and Afterglow Observations}

\author[0000-0001-5939-644X]{P. Nuessle}
 \affiliation{Department of Physics,
   The George Washington University,
   725 21st St. NW, Washington, DC 20052}
 \affiliation{Astrophysics Science Division, NASA Goddard Space Flight Center,
   8800 Greenbelt Rd, Greenbelt, MD 20771}
 \author[0000-0002-4744-9898]{J. L. Racusin}
 \affiliation{Astrophysics Science Division, NASA Goddard Space Flight Center,
 8800 Greenbelt Rd, Greenbelt, MD 20771}
 \author[0000-0002-9408-3964]{N.E. White}
 \affiliation{Department of Physics,
   The George Washington University,
   725 21st St. NW, Washington, DC 20052}


\correspondingauthor{P. Nuessle}
\email{nuessle167@gwu.edu, nnuessle@nasa.gov}



\begin{abstract}

Using an established classification technique, we leverage standard observations and analyses to predict the progenitors of gamma-ray bursts (GRBs). This technique, utilizing support vector machine (SVM) statistics, provides a more nuanced prediction than the previous two-component Gaussian mixture in duration of the prompt gamma-ray emission. Based on further covariance testing from \textit{Fermi}-GBM, \textit{Swift}-BAT, and \textit{Swift}-XRT data, we find that our classification based only on prompt emission properties gives perspective on the recent evidence that mergers and collapsars exist in both “long” and “short” GRB populations.

\end{abstract}



\section{Introduction} \label{sec:intro}
Until recently, GRB data and theory generally supported a Gaussian mixture of GRB properties corresponding to two classes due to massive stellar collapse of hydrogen- and helium-stripped stars, either triggered by supernova or collision \citep[collapsars,][]{Wolf-Rayet_insanity, Wolf-Rayet_LTE, hydrogen_evelope, detach_those_binaries} and compact mergers including at least one neutron star \citep[mergers,][]{also_a_two_classes_paper, 2022Galax..10...77S, 2022A&A...657A..13T, progenitor_proof}. The prompt gamma-ray emission periods from mergers are generally shorter than some instrument-dependent duration boundary, (generally estimated $\sim$2 seconds) while those from collapsars are longer \citep{Real_two_classes_paper}. The collapsar mechanism was first confirmed by the association of SN 1998bw, a Type Ic supernova, with GRB 980425, a long burst \citep{real_980425A, GRB980425}. Since then, dozens of supernova-associated long GRBs have been discovered \citep{three_supernovae, supernova_paper, GRB_030329, GRB_060218}. The merger hypothesis was originally supported by the discovery of short bursts like GRB 050509B, which lay on the outskirts of old elliptical galaxies \citep{more_GRB050509B, GRB050509B}. This progenitor scenario was confirmed by the coincident detections of GW170817/GRB 170817A/AT 2017gfo, a binary neutron star merger and the association of an r-process kilonova \citep{GW170817_paper,GRB170817A}. GRB 170817A spurred a search for more archival kilonova signatures following bursts, and some were found \citep{GCN_230307A, GCN_070809, GCN_130603B, GCN_211211A, GCN_160821B, GCN_150101B, GCN_111005A, GCN_060614}. However, there are now directly observed classifications where kilonova signatures followed long-duration bursts \citep[long mergers,][]{GCN_230307A, new_230307A_paper, yet_another_211211A, GCN_150101B, GCN_211211A, long_060614} and where one supernova signature followed a short-duration bursts \citep[short collapsars,][]{short_collapsar_paper, short_collapsar_other_source}.

 We used these unexpected associations combined with the more familiar long GRB-supernovae and short GRB-kilonovae to train our new classifier so that it would not be wholly based only on the duration or hardness of the burst, but rather trained on known associations via this list. The formation and transformation of the progenitor system should cause profound observational effects on the resulting GRB, making these "muddled" bursts separable based on other factors. A compact binary system is more likely to have migrated out of its star-forming region during the long period between when the stars collapse to form compact objects and when they merge, leaving the binary with little to no surrounding gas. This causes the afterglow to be suppressed, as it is the interactions with the environment that create broadband afterglow emission \citep{Gehrels_original, kilonovae_overview}. This just one of many features we could use to identify the progenitor system.
It is unlikely that the two burst types have different emission mechanisms given that they appear to have similar relationships between their observed flux; luminosity; and peak, isotropic, and collimation-corrected energies, but it is clear that higher peak energies are generally associated with merger events \citep{same_emission_mechanism, definitely_same_emission_mech}.

The motivation to build one or more burst classifiers came directly from these exceptions to the paradigm which assumed that collapsars were long and mergers were short. Some recent methods have focused on the short kilonovae alone or on searching for third (or fourth) classes of GRBs to explain the property overlap \citep{Dimple_Two_Merger_Types, Three_Classes_paper, also_a_two_classes_paper}. We instead examine $E_{p, prompt}S_{\gamma, \text{prompt}}^{-1}$ vs. $\text{T}_\text{90}$ as in \cite{Goldstein_New_Classifier} for discriminatory power \citep{yet_another_prompt_ratio}. We find evidence for a continuum between the two progenitor classes of GRBs.

In Sec. \ref{Data}, we define our known-progenitor training sample and the rest of the sample selection. 
In Sec. \ref{methodology}, we outline the support vector machine (SVM) machine learning tools used in this paper. 
In Sec. \ref{analysis}, we determine the best discriminator and used those factors to create a probabilistic SVM model and attempt to quantify its covariates and accuracy. In Sec. \ref{Discussion}, we discuss our results and suggest where further work may be necessary. In Sec. \ref{conclude}, we conclude.

The Hubble constant is assumed to be $H_0$=70 km s$^{-1}$ Mpc$^{-1}$. The significance threshold is $\alpha=0.01$, which is approximately $2.32\sigma$ \footnote{\url{https://www.scribbr.com/statistics/statistical-significance/}}.

\section{Sample Selection} \label{Data} 
Prompt emission data were utilized from the \textit{Fermi}-Gamma Ray Burst Monitor (GBM) GRB catalog from 2017 August through 2023 May \citep{Fermi_Ten-Year_Catalog}. This instrument was chosen due to its ability to characterise the prompt emission energy spectrum over a wide bandpass and its large uniform sample data set. We chose to use the X-ray afterglow (where available) to study potential selection effects. This afterglow sample runs through the same period of time and came from the \textit{Neil Gehrels \textit{Swift} Observatory} X-Ray Telescope (XRT) \citep{Swift_Fit_Algorithm}. We included this data as a check on our support vector machine for the two progenitors, as characterization of the early afterglow, including the important plateau period, may be related to continued energy injection into the circumburst environment \citep{Gehrels_original, Judy_Efficiency}. These samples are cross-associated through the trigger times as matched through the \textit{Swift}-Burst Alert Telescope (BAT) GRB catalogs. When both published GRB catalogs and preliminary characterization are included, all three data sets are complete through 2023 May \footnote{\url{https://swift.gsfc.nasa.gov/results/batgrbcat/\#Contact}, \url{https://swift.gsfc.nasa.gov/archive/grb_table/}}. We refer to this as the GBM-BAT-XRT sample. If more XRT data were needed or we wanted to perform our own data analysis directly on the \textit{Swift} burst data, it was retrieved using the \texttt{swifttools} python package as a burst analyzer \footnote{\url{https://www.swift.ac.uk/API/ukssdc/data/GRB.md}}. If more detailed GBM data was required, it was downloaded from the GBM archive and processed using the \texttt{gbm-data-tools} python package \citep{gbm-data-tools-code}.

If only the prompt emission was needed, for instance when tentatively predicting the progenitor of all bursts in the GBM GRB catalog, only the GBM sample was used. However, when testing for distance limitations on the classifier, a redshift-associated GBM sample combined with a subsample of the overlapped GBM-BAT-XRT sample that has a measured redshift was used. We also used an overlapped GBM-BAT-XRT sample with afterglow associations as a stand-in test for the GRB environments (and therefore at least some burst progenitors). These redshifts were found using the BAT redshift table and J. Greiner's GRB table \footnote{\url{https://swift.gsfc.nasa.gov/results/batgrbcat/\#Contact}, \url{https://www.mpe.mpg.de/~jcg/grbgen.html}}.

We refer duration to mean $\text{T}_{\text{90}}$ the time between which 5 and 95\% of a burst's fluence over 50 to 300 keV ($S_{\gamma, \text{prompt}}$) is observed. All bursts referred to as long in this work are taken to have a prompt emission duration longer than 4.2 seconds in \cite{Fermi_Ten-Year_Catalog}, and duration shorter than 4.2 seconds as "short". Many authors (e.g. \cite{new_230307A_paper}) now recognize that the duration separation between a "long" and a "short" burst may not always be two seconds as determined for the Compton Gamma-ray Observatory (CGRO) Burst and Transient Spectrometer Experiment (BATSE) sample \citep{BATSE_inst_paper, Two_Classes_Paper, prob_better_BATSE_paper}. Instead, it may depend on the instrument, its trigger criteria, energy range, and sensitivity as hypothesized by \cite{instrument_dependent_classes}. It is also known to depend on the burst's redshift, as time dilation affects the duration, fluence, and counts detected above background. These authors found that the $\text{T}_\text{90}$ at which a GRB is equally likely to be a collapsar or merger is 0.8, 1.7, and 3.1 seconds in the BAT, the GBM, and BATSE data, respectively \citep{swift_inst_paper, Fermi_Ten-Year_Catalog, Swift_Catalog, fermi_inst_paper}. Taking this under advisement, we used 4.2 seconds, which the authors of the Fourth Fermi-GBM GRB Catalog fitted using a lognormal fit to the $\text{T}_\text{90}$ distribution \citep{Fermi_Ten-Year_Catalog}.
\subsection{Detected Progenitors} \label{Progenitors}
In this section, we describe how we selected our detected progenitors. Each class is treated separately.
\subsubsection{Collapsar Selection}
The presence of a localization-consistent supernova or late afterglow optical light curve bump typical of a supernova is taken to represent a collapsar origin for the gamma-ray burst. We used these characteristics to compile a list of 42 typical "long collapsars", 1 "short collapsar" (GRB 200826A), and one notable "exotic burst" which seemed to have features of both progenitor types (GRB 210704A) by searching the literature and the GCN Circulars \footnote{\url{https://gcn.nasa.gov/circulars}}. The large number of supernovae can be attributed to their relative brightness and extensive literature of detection.
The vast majority of collapsar bursts were taken from in Table 7 of \cite{supernova_paper}. The four exceptions are GRB 221009A \citep{GCN_221009A}; GRB 211023A, \citep{GCN_211023A}; GRB 150210A\citep{2023MNRAS.518.6243J}; and GRB 200826A, which is the only short collapsar \citep{short_collapsar_paper}. In total, there were 43 bursts with associated supernovae used to train our classifier our sample (Table \ref{Data_Table}).

\subsubsection{Merger Selection}
The presence of an optical or infrared kilonova signature is taken to represent a neutron star merger origin, which can be observed with either distinct spectral features or a weak bump in the infrared light curve.  We also select short GRBs in the outskirts of their host galaxies as indicators of a possible merger. There are fewer known merger events and their associations can be more preliminary. Each of our selected merger associations has a unique associated paper or GCN Circular (Table \ref{kilonova_table}). Nine of those have direct kilonovae detections, meaning that a kilonova model normalized by AT2017gfo could be fit to their infrared or optical data. These are the progenitors of which we are most certain. The three bursts fit with a weak kilonova may only have a bump in the IR associated with r-process nucleosynthesis. Along with the burst with a short spectral lag \citep[another common feature of short GRBs, see:][]{Alyson_thesis, Spectral_lags_be_working}, these are the bursts we have some confidence of being mergers. Finally, the eight position inferred mergers are those of which we are least certain.

To increase the sample of known mergers, we included events from \cite{Galactic_Mergers_Paper}, who performed a detailed study of the host galaxies associated with short GRBs. This has been proven in previous analyses such as \cite{galactic_sorting} and \cite{galactic_classifier_2} to meaningfully distinguish between two populations of gamma-ray bursts in the Greiner catalog. These authors similarly found that no combination of factors may cleanly predict a burst's progenitor, and the primary relationship they find is redshift-dependent \citep[p. 28][]{galactic_sorting}. Their choices of effective amplitude (scaling down the signal to noise ratio of the peak flux to the background to make the duration less than two seconds) and brightness fraction (another measure of how active the star formation in the GRB's area of the galaxy is) rather than fluence and peak energy are also themselves redshift-dependent. The calculations to transform the first through different frames of reference are a corollary of the calculations in Sec. \ref{analysis}. However, when they use a Naive Bayes classifier trained on the Greiner catalog definition of Type I and II GRBs, they remove this dependence in favor of a high-dimensional model in the prompt emission and galactic information \citep[p. 6,][]{galactic_classifier_2}. We chose to instead simplify our model to three factors so we could train it on the few known progenitors. From Ch. 14.9 of \cite{get_to_cite_a_book_for_once}, we also assume that the majority of merger GRBs occur at the outskirts of their host galaxies due to the large kick velocities from one or both of the supernovae the formed the neutron stars. We therefore assumed that short GRBs found more than a certain distance away from the center of their host were likely mergers. For each detected host galaxy, the authors performed positional and spectral analyses between the GRB and a catalog of galaxies to determine a probability of chance coincidence between the GRB afterglow and the host galaxy itself. The percent chance of coincident association between the GRB and its host galaxy were low for our sample but ranged up to 5\% instead of 1\%. We assume that if a GRB was more than 1.96 radii away from the center of its host galaxy, it was more likely to be in a region of inactive star formation \citep{Sylvain_why_so_long}. This value was chosen because if the radial stellar light profile of a galaxy were Gaussian, then 95\% of the stellar mass density would be within this radius. While the mass density of a galaxy is best represented by a Navarro–Frenk–White profile, \cite{galactic_lightcurves} found that a constant stellar mass over luminosity ratio for an exponential radial luminosity profile would be a good zeroth-order approximation to the real kinematics. This also means the luminosity profile would not be too skewed towards the edges, meaning that a Gaussian would be a good approximation. As a result, we could not test many GRBs, as the effective radius of their host galaxy is unmeasured. As for the rest, it was unclear if they simply had not moved from their area of formation or if they were true short collapsars. We chose not to take them as mergers them without more information on their origin. This netted us 8 further short-merger type GRBs.

\section{Methods}\label{methodology}
The goal of this classification is to differentiate between the two accepted progenitor classes of GRBs. As there are a large number of possible input variables but only a small amount of training objects (and two output classes), simpler methods and statistical models are mathematically preferred. To that end, we explored multiple machine learning methods before settling on a support vector machine \citep[SVM,][]{proper_SVM, basics_of_SVMs}. For completeness, we include some of them here as possibilities for future work.
The primary classification method used in this paper is SVM as it has many useful features in our case, as well as being a tested method in astrophysics \citep{SVM_and_boosted_trees, Bat_Trigger_SVM, GW_SVM_evidence, More_SVM_evidence}. An SVM can be understood as a classification based upon the creation of a ($n-1$)-dimensional hyperplane in the $n$-dimensional geometry created by the $n$ features of the objects to be classified \citep{basics_of_SVMs}. This plane is defined by a kernel function that determines its shape and properties--we used a radial function as the kernel to approximate the creation of a neural network without requiring a large training set. SVM is ideal for a large data set with a small training set, as it minimizes the training set as part of its algorithm \citep{basics_of_SVMs}. Per \cite{Pharmacology_paper}, it acts consistently over many sizes of training sets. The GRBs with known classification are then chosen to be the training data. Those members of the training set that lie closest to the hyperplane become its "support vectors," which dictate the hyperplane's angle, curvature, and placement. SVM is not a precise method, and it has no inherent classification confidence, though there are affiliated methods such as Platt scaling and the relevance vector machine which make this possible, at least in the binary case \citep{variational_RVM, Platt_Scaling}. We chose this method with the application of Platt scaling because in the binary case, it requires the fewest assumptions while approximating more complex algorithms.

\begin{figure}
        \includegraphics[width=0.35\textwidth]{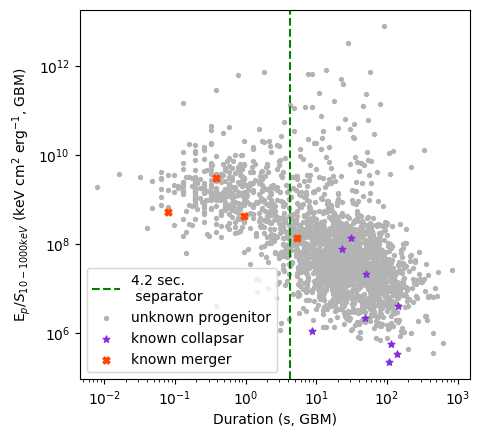}
        \label{fig: Goldstein unfitted}
        \includegraphics[width=0.42\textwidth]{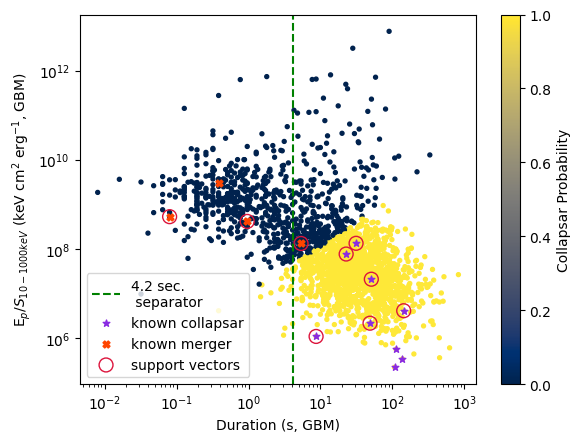}
        \label{fig: Goldstein Fitted bordered}
        \includegraphics[width=0.42\textwidth]{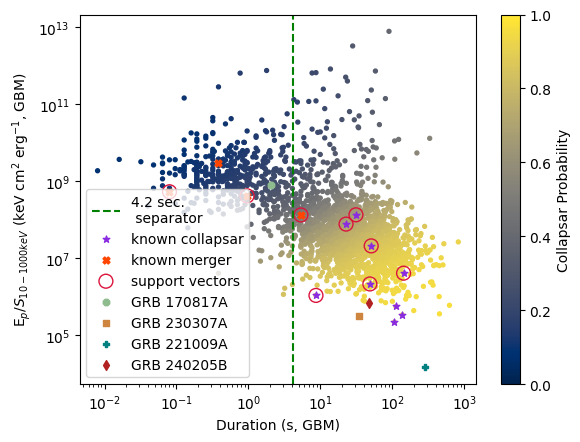}
        \label{fig: Goldstein Fitted}
    \caption{The prompt energy ratio ($E_{p, prompt}S_{\gamma, \text{prompt}}^{-1}$) is strongly correlated with $T_{90}$, is a good separator of events with know progenitors, making it the basis of our classification.  The 4.2 second long/short duration demarcator derived by the Fourth GBM GRB catalog as a dotted green line \citep{Fermi_Ten-Year_Catalog} demonstrates how duration alone is an insufficienct classifier. {\it Top}: The unknown long and short bursts are unsorted. {\it Middle}: Our SVM binary classification is applied to the sample. {\it Bottom}: We apply Platt scaling to the classifier to make it probabilistic rather than binary, and  highlight other notable bursts with known progenitors. As the error bars make it difficult to distinguish individual events, a version of this plot with error bars is in App. \ref{Error Bars}.}
    \label{fig: Goldstein All Data}
\end{figure}

Other classification methods were also explored including random forest, which shows promise as a future method when the training sample is significantly larger. In this algorithm, thousands of decision trees are created and evaluated based on how many groups they split the data into \citep{random_RF_paper}. The "deeper" a tree is, (i.e. the more decisions it includes) the more subtrees it includes, and the less each individual deep tree is weighted in the result \citep{random_forest}. This means that the algorithm can suffer from overfitting if the trees are allowed to become too deep or there are too many. However, boosting the classifier by making it a weighted linear combination of multiple algorithms makes it more robust to this error \citep{SVM_and_boosted_trees}. Random forest is also computationally intensive when calculating a large number of trees, and it is strongly suggested that the training sample be a significant fraction of the test sample, as the training sample is used to bootstrap smaller "sub-training" samples for each tree to ensure that they are more random than the overall model \citep{random_forest}. We chose to discuss it even though we did not have enough training data to properly utilize its power in testing. As more known progenitors are found, it may be possible to use random forest or gradient boosting trees to create more complex models than the one presented in this paper. Some authors such as \cite{I_was_right_I_read_this_one}, are already using it. They use the Greiner table with some interpretation to distinguish between Type I and Type II GRBs, rather than directly training on known progenitors. This allows them enough data to tune the model. Our relative lack of training data also limited our ability to use neural networks.

\section{Analysis}\label{analysis}

\cite{Amati_original} observed a correlation between $E_{p, prompt}$ and $E_{\gamma,iso}$ using rest-frame prompt spectral properties of GRBs. Updates to the model later proposed that long and short GRBs might follow different linear relationships \citep{updated_Amati}.  There is debate in the literature if the so-called Amati relation is due to intrinsic or observational selection effects \citep{Amati_Broken, Amati_instrument_dep}, though some review articles take it to be at least partially intrinsic to the system \citep{I_dont_want_to_wade_into_this_fight}. 
However, when taken to the observer frame, \cite{Goldstein_New_Classifier} found that the Amati relation in BATSE data had a suggestive relationship to features also known to partially differentiate between progenitors such as burst duration and hardness ratio. We take this "prompt energy ratio" feature to be $E_{p, prompt}S_{\gamma, \text{prompt}}^{-1}$. 
After observing the progenitor classes' visual separation in Fig. \ref{fig: Goldstein All Data}, we created an SVM based on the prompt emission variables $E_{p, prompt}S_{\gamma, \text{prompt}}^{-1}$ and duration for easier visualization in two dimensions. In three dimensions, it can be more easily manipulated as $E_{p, prompt}$, $S_{\gamma, \text{prompt}}$, and duration.

\begin{figure}
    \centering
    \includegraphics[width=0.46\textwidth]{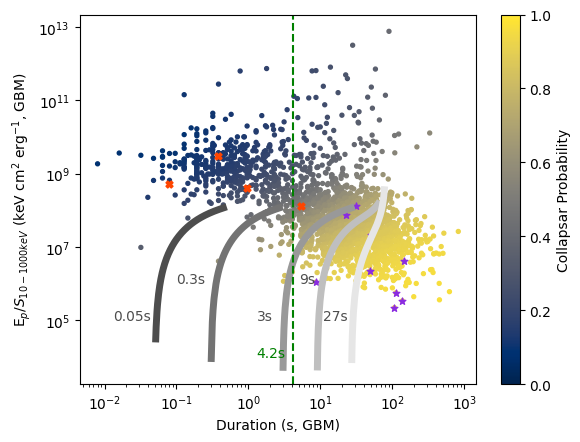}
    \caption{Each of the greyscale curves is a different simulated burst with the same spectral model, but a different $\text{T}_\text{90}$ in the rest frame. We simulated their observed properties from $0.01 \leq \text{z} \leq 8$. This simulation demonstrates that distance alone does not explain all observed progenitor relationships in Fig. \ref{fig: Goldstein All Data} nor most of the uncertainty. It does so as the majority of the simulated data is found to run perpendicular to the relationship, rather than along it. The spectral values chosen ($E_p$=150 keV, $E_{\gamma, iso}=3\times 10^{52}$ ergs, and modeled using a Band spectrum with $\alpha=-0.8$ and $\beta=-2.76$) were the median GBM catalog values for short bursts fitted with a Band or SBPL function.}
    \label{fig:Multiburst_distance_GRB_model}
\end{figure}

The advantage of the prompt energy ratio-$\text{T}_\text{90}$ correlation is that it includes many bursts from both progenitor classes because the selection criteria are very broad and only come from the observed-frame prompt emission. The model is difficult to interpret physically but is strongly related to the Amati and Ghirlanda relations \citep{Goldstein_New_Classifier}. However, both relations may be instrument-dependent \citep{Amati_Broken, Amati_instrument_dep, yet_another_prompt_ratio}, making it possible that the proposed classifier should be retrained each time it is applied to an instrument with a different bandpass or spectral fit. As derived by \cite{Im_so_tired}, $E_{p, prompt}^2/S$ should be weakly related to redshift, z. \cite{Goldstein_New_Classifier} then simplifies this ratio to $E_{p, prompt} S_{\gamma, \text{prompt}}^{-1}$, finding that it creates a Gaussian mixture in BATSE bursts. When we then plot this ratio for GBM bursts against the $\text{T}_{\text{90}}$, we see that there is a two-dimensional Gaussian mixture in the duration and spectral features of the burst. A successful classifier should be able to identify if an object has had a merger or collapsar progenitor, regardless of the burst's duration, on both the training and test data sets. Therefore, it appears suggestive that this distribution is mixed even within the same duration.

In searching all GBM bursts (Section \ref{Data}), we find that 13 of the 2310 bursts with a measured peak energy also have a known progenitor (Fig. \ref{fig: Goldstein All Data}) (9 known collapsar systems and 4 known merger systems). We used these 13 known progenitors to create an SVM classifier based on the prompt energy ratio and the $\text{T}_{\text{90}}$ of these bursts. The full SVM model is available for use and transformation as needed \footnote{10.5281/zenodo.11107782}. The hyperplane is a line segment (a 1-D hyperplane) that initially appears mostly straight, though strongly tilted according to a burst's peak energy, based on 6 support vectors of the 13 known progenitors. We then applied this classification to all bursts in the test data to extract interesting features. (Due to the error bars' relative size, a version of the plots including them can be found in App. \ref{Error Bars}.) We find that the border between merger and collapsar bursts is tilted in the peak energy-time axis and slightly above the 4.2s divider given in \cite{Fermi_Ten-Year_Catalog}.

We examined several potential observational selection effects, including the real distribution of distance, simulated scaling of distance effects on other parameters, and selection effects on prompt and afterglow fluence, that could be covariates for our classifier--that is, secondary factors that could influence or better predict the progenitor of a GRB. After noticing that many papers that had shown a strong redshift-dependence in their correlations used redshift-dependent variables, \citep[including ours,][]{galactic_sorting} we chose to simulate if a single long or short burst with or without error was simulated at different redshifts could reproduce our correlation. Instead, it predicted a different burst property distribution than the one found in the GBM catalog. These bursts were simulated using $E_{p}=350$ keV, isotropic energy ($3 \times 10^{52}$ ergs), and median values of the Band spectral indices of bursts in the GBM catalog shorter than 4.2 seconds best fit by either a Band or SBPL function ($-0.800, -2.76$). From that and duration at a redshift of zero of 0.05, 0.3, 3, 9, and 27 seconds, we were able to simulate the peak energy, fluence, and observed duration of each burst at a list of randomized distances between z=0.01 and 8. These values were simulated as:
for $z_{sim} \in [0.01, 8]$
\begin{equation}
E_{p}=\frac{E_{p,0}}{1+z_{sim}}
\end{equation}
\begin{equation}
\begin{split}
\text{T}_\text{90} & =\text{T}_{\text{90},0} \times \int_{t_a}^{t_b} N(0, \text{T}_{\text{90},0}(1+z_{sim}))- \\
& 0.4 \frac{e^{-erfinv(\times\text{T}_\text{90}^2) \text{T}_\text{90}^4}}{\text{T}_\text{90}} dt
\end{split}
\end{equation}
where
\begin{equation}
N(0, \text{T}_{\text{90},0}(1+z_{sim}))|_{t_a}=-100
\end{equation}
\begin{equation}
N(0, \text{T}_{\text{90},0}(1+z_{sim}))|_{t_b}=100
\end{equation}
\begin{equation}
\begin{split}
S_{\gamma, \text{prompt}} & =\frac{E_{\gamma, iso, 0} (1+z_{sim})}{(4\pi k d_{lum}^2)} \times \int_{t_a}^{t_b} N(0, \text{T}_{\text{90},0} (1+z_{sim}))- \\
& 0.4 \frac{e^{-erfinv(0.9 \text{T}_\text{90}^2) \text{T}_\text{90}^4}}{\text{T}_\text{90}} dt
\end{split}
\end{equation}
Where
$d_{lum}$ is the luminosity distance in centimeters of an object at $z_{sim}$, 0.4 is the normalization of the normal distribution, and k is the k-correction of a Band function--that is, the change in fluence due to spectral changes with redshift and instrument sensitivity. We also extrapolate the fluence from our received range of 10 to 100 keV to 1 to 10000 keV:
\begin{equation}
k =\frac{\int_{10\ keV}^{1000\ keV}f(E, E_p=350 keV, \alpha=-0.8, \beta=-2.76) dE}{\int_{\frac{1\ keV}{1+z_{sim}}}^{\frac{10000\ keV}{1+z_{sim}}}f(E, 350, -0.8, -2.76) dE}
\end{equation}
where $f(E, E_{p}, \alpha, \beta)$ is the Band spectral function as laid out in \cite{Band_Function_Paper}. The peak energy, ($E_p$) low energy power law index, ($\alpha$) and high energy power law index ($\beta$) values were selected as the averages of \textit{Fermi}-GBM bursts shorter than five seconds and best fit by Band functions:
\begin{equation}
    f(E) = 
\left\{
    \begin{array}{lr}
        \left(\frac{E}{100}\right)^{\beta} e^{\beta-\alpha} \left(\frac{(\alpha-\beta)E_{p}}{100\alpha} \right)^{\alpha-\beta},
        \text{ if } E \leq \frac{(\alpha-\beta)E_{p}}{\alpha}\\
        \frac{E}{100}^{\alpha} e^{-\frac{\alpha*E}{break_E}},\text{ if } E > \frac{(\alpha-\beta)E_{p}}{\alpha}
    \end{array}
\right\}
\end{equation}
Rather than resembling the real burst distribution, these simulations created a distinct distribution roughly perpendicular to the observed data, as can be seen in Fig. \ref{fig:Multiburst_distance_GRB_model}. These simulations appear to predict that distance explains some of the scatter in bursts with similar properties rather than the full distribution. 

As another test to see if redshift dominates the correlation, we split our sample in two by high and low redshift, here defined to be 1.5, roughly both the median and mean of the redshift sample, and found that they showed no significant difference in the way they were correlated at a significance level of 0.01. This and all further tests on the difference between two populations were performed using a Student-T test with a significance threshold of 0.01. All tests on the similarity of two populations were done using a power analysis with effect size 0.5 (meaning we expect a moderate different in the populations) and power 0.99 (meaning the analysis will detect that moderate effect with 99\% accuracy if there is one). We examined the observed redshift and the simulated distance separately as redshift would indicate if our classification showed limitations dependent on distance, whereas distance simulations would indicate if distance was a significant predictor of the classifier. The sample with redshift could not be rejected as representative of the entire GRB sample at a significance level of 0.01, though it also could not be accepted as representative of the entire sample at that level. We postulate that this classification is not well explained by observed redshift or simulated distance alone, but rather by multiple physical factors.

Next, we calculated an afterglow fluence, as it is considered a secondary test of the progenitor system. We did this by assuming that the integrated X-ray afterglow flux (i.e. fluence) over 0.3-10 keV in \textit{Swift}-XRT during the "plateau" phase of the early afterglow would be related to the environment of a GRB, and that the environments of merger and collapsar-type bursts would be relatively distinct. We used the following to calculate this fluence:
\begin{equation}
S_{\text{X}, AG}=Q*\int_{t_{start}}^{t_{stop}} g(t, E_{i}, \alpha_{i}) dt
\end{equation}
Where $Q$ is a normalization factor found by setting the function equal to the afterglow flux at 11 hours, $g(t, E_{i}, \alpha_{i})$ is a power law in time dependent on the number of breaks present in the afterglow lightcurve, and the start and end times are measured from the XRT lightcurve. We found that when we split our full afterglow sample in two by high and low afterglow fluence at the our calculated median value of $6.6 \times 10^{-7} \text{ergs cm}^{-2}$, the data show only a slight difference in the way they were correlated at a significance level of 0.01. 
The duration and prompt energy ratio of this subsample was found to differ from the main sample at a significance level of 0.01, so we must reject the null hypothesis that these bursts represent the underlying distribution. Instead, we hypothesize that their afterglows are brighter than average, meaning there are selection effects on which bursts have detectable afterglows. This would not negate the fact that this test predicts two slightly different afterglow (and therefore environment) populations that are classified slightly differently. It rather implies that it only does so on a small fraction of the data. We therefore assert that the classifier appears to differentiate based on a burst's environment as a result of progenitor classification on at least some data.

As a test of the limits of our classifier, we split the sample into high and low prompt fluence subsamples across the GBM catalog median value of $3\times 10^{-6}\text{ erg cm}^{-2}$. This fluence was measured at 10-1000 keV over the burst duration, which we take to be time between the start and end time of the fluence spectrum. The $\text{T}_\text{90}$ is not used as it is related to the fluence over 50-300 keV. We find cuts across this boundary lead to significantly different classification fits and distributions. Furthermore, one of these distributions is significantly different from the main distribution; both correlations are significantly different from each other. We take this to indicate that the classifier strongly depends on a burst's prompt fluence. Furthermore, despite having roughly equal numbers of "high" and "low" fluence bursts, it appears that the classification distribution as a whole is more similar to the low fluence bursts. Meanwhile, the classification is very different across all three groups. This could mean that the classifier works better on high-fluence bursts, as it changes so much when they are subtracted from the distribution.
\begin{figure}
        \includegraphics[width=0.46\textwidth]{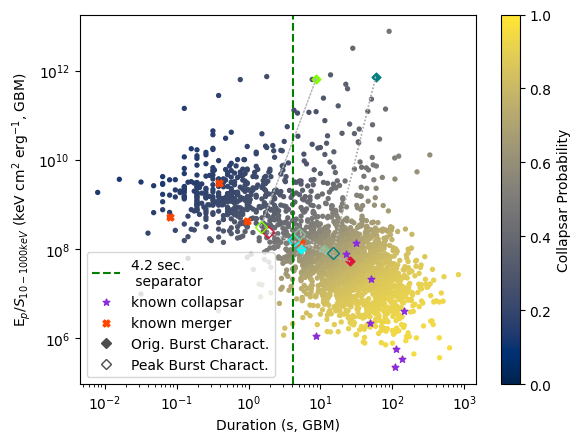}
        \label{fig: Goldstein Fitted Special}
        \caption{When performing time-resolved spectral fits classifying only the short hard spike, and repeating their parametrizations, the burst observations moved within the classification parameter space. This is a source of uncertainty for many of bursts both near and distant from the classification border, and it is worth deeper study.}
\end{figure}
Due to the small number of known progenitors, we also choose to test selection effects within those. To do so, we randomize the number of collapsars that "exist" in the sample between 1 (to make sure two classes would be present) and 18 (twice the number that are present in the original sample). We do not randomize the mergers as 3 of the 4 of them are used as support vectors in the classifier. We then select that many collapsars randomly with replacement from the original collapsar sample. Each "new collapsar" is then assigned a random the prompt energy-fluence ratio and the burst duration using a Gaussian with mean 0 and standard deviation of the uncertainty of the measurement (0.15 and 0.12, respectively). These two steps would approximate the residual bootstrap of a linear/nonlinear model, where the data sample and residuals are resampled with replacement, then assigned randomly. 
This simulation indicated that our previous analysis using assumptions for an t-distributed variable was limited at best, as the distributions of the test variables were not found to be normal. However, we did find that the percentage of collapsars, and the classification percentages of GRBs 170817A and 230307A were near the peak of their distributions. This indicates that the classifier is possibly limited by its statistical method and physical model, rather than the sample size. This was predicted by our choice of classifier (which would minimize the number of used progenitors) as well as work performed by \cite{Pharmacology_paper}. When more progenitors become available, it should be possible to revisit this analysis using a more self-correcting method like gradient boosting trees, as we mentioned in Sec. \ref{methodology}.
As changing the known progenitors of the classifier would either have no effect (if we did not change a support vector) or a profound effect, we chose to test the meaning of each classification more indirectly. We instead examined five GRBs in different areas of interest on the plot (Fig. \ref{fig: Goldstein Fitted Special}). Two (GRBs 081109 and 111010B) were high indicator variable outliers. When we split them by features in their light curves and performed time-resolved spectral analyses and remeasured their durations, we found that the listed peak energies in the catalog were poorly constrained because they were originally fit by a simple power law. When the classifier selected a secondary fit to use its peak energy, the derived value was not properly constrained, existing at an upper limit. We therefore conclude that bursts best fit by a power law are sometimes poor candidates for this method. However, when all bursts best fit by a power law are removed from the model, all but one merger are removed as well. From here forward, we used the second most likely fit. 

Two GRBs (110820C and 100907A) lie near the center of the distribution, one with a slight classification preference and one without (36\% and 45\% predicted similarity to a merger, respectively). When only what appeared to be their main emission was selected, refit, and reclassified, their progenitor predictions become less certain (now 53\% and 56\% predicted similarity, respectively). Finally, GRB 120308B was chosen as it was a typical long collapsar. However, when its spectrum was refit on a shortened duration around the first pulse of the burst, it became slightly more similar to a merger than a collapsar.

We also included the known progenitor bursts 221009A and 230307A as checks on the method (see Fig. \ref{fig: Goldstein All Data}). They appear on the far bottom right of the distribution with most estimations of the peak energy, as spectral fits published by teams other than the GBM team for GRB 230307A are fit with an evolving 2SPBL fit and that of GRB 221009A with an evolving Band plus power law component \citep{230307A_spec, 221009A_spec}. Both are classified as long collapsars, where one is a known long collapsar and the other a long merger, which we suspect to be due to their extraordinary fluence. However, we also cannot rule out that there is separation line evolution with fluence due to its relationship with $\text{T}_\text{90}$ as seen in our prompt fluence separation test. It is also probable that any classification based on prompt emission alone would be limited, as the central engine and environment are needed to fully determine information about a burst's progenitor. We include these progenitors as a indicator that further refinement on this classifier is a topic for future work, including our maintained GitHub. As a field, GRB science is frequently surprised by energetic and duration outliers such as these. They may indicate a gap in understanding that 
future observations or better analysis and machine learning techniques may help to resolve. These instruments may help us to detect more of these objects, increasing our knowledge of which ones are outliers and which represent new physics, as well as to coordinate observations of their multiwavelength components with other observatories.
This parameterization of the prompt emission still shows promise as a classifier between two burst classes as seen in Fig. \ref{fig: Goldstein All Data}.
We do not find any evidence of accessory classes in line with the results of previous machine learning studies \citep{2022Ap&SS.367...39B, 2022Galax..10...77S, 2022A&A...657A..13T}.
\section{Discussion}\label{Discussion}

Considering the results of our analysis of our prompt energy ratio vs. $\text{T}_\text{90}$ model, we find that the SVM classifier successfully predicts GRB progenitors with some high-fluence and spectral model limitations. We find no probabilistic reason to split our burst distribution into more than two types based on the features. Our main motivator for this assertion is the lack of a third class of progenitor system, though the return of a Gaussian mixture in three dimensions also contributes. There is also the fact that the classifier correctly classifies all its training data. The only exceptions to its correct classification of known progenitors was 230307A, which we believe to have exceeded some as yet unknown brightness or axial limitation. In this situation, a simpler model should be favored, here meaning one with fewer classes.

We find that the $\text{T}_{\text{90}}$ cutoff between collapsar and merger bursts is not only slightly above 4.2s but also dependent on the prompt energy ratio of the burst. This is better centralized in our tests than having it depend on the hardness ratio of the burst. It indicates that there are some MeV-peak bursts still characterized as merger-type, and some middle-fluence bursts which are undoubtedly collapsars. We are unsure of the cause of these, but we did try to model them using our known progenitors list (Table \ref{progenitor table}). Many of those progenitors never made it into this or any other sample as they were not seen by GBM or BAT or XRT. We can only speculate that these shortened collapsars may be either burst precursors or have weak jets \citep{200826A_weak_jet, 200826A_geometry}, the long mergers may have gone through a rotationally-supported hypermassive neutron star phase \citep{Sigh_Rowlinson_Paper}, or the exotic burst could be anything from a white-dwarf-neutron star merger to a particularly unusual supernova \citep{exotic-burst-source}. In fact, papers numerically modelling short and intermediate-length engines for collapsars such as \cite{short_int_collapsar_engines} have suggested that their jet lightcurves may be strongly biased towards the precursor jet phase and highly variable. With the data as taken, it is difficult to determine which interpretations show promise. However, we did create this classifier with the ability to return a probabilistic classification, as seen in Table \ref{fig: unnecessarily long table}. Given that most bursts have a fairly certain classification, we interpret that most GRBs fall squarely into our known progenitor systems of either a high-mass stripped star or a colliding neutron star binary per the predictions of \cite{Real_two_classes_paper}. As more "short" collapsars are added to this model, it may become easier to determine if it is a physically meaningful classifier given that the line of determination is neither straight nor denoted in the unlabelled graph. We could also try adding features that characterize the environment, variability timescale, and spectral lag which might refine our models. The two paths to accomplish this will be adding more known progenitors and creating physical models that include more variables. While SVM minimizes the number of training data points it uses to build the model, it is very likely that the new data will include one or many points that are more significant than the ones currently present. It is also highly likely that when we use models that predict a progenitor's behavior in more than two (or three) dimensions, the SVM itself will separate the bursts differently as the higher dimension plane will have a different minimal path. We believe these two paths serve as a path forward, and prove this distribution to be a first step rather than a finalize model.

In the interest of collecting more data about merger environments, we recommend that broadband observations of long mergers similar to GRB 211211A be undertaken. Papers that performed similar work would include \cite{AT_2017gfo_SED, 221009_SED, Alex_and_his_undergrad}. This burst is close enough that the rapidly fading afterglow X-ray flux was constrained at late time \citep{New_GCN_211211A}. We suspect in addition that the burst's environment may be unusually dense, making it a good candidate for radio follow-up. If a radio afterglow signal is detectable, it is possible that the GRB environment was relatively dense, changing our understanding of where long mergers evolve. However, if the afterglow was difficult to observe in the radio, it may be possible that long mergers are instead a result of time-dilation of the duration. We do not believe this to have a strong effect on the classifier, but rather recommend it to be investigated as a possible cause of GRB duration lengthening.

Our classifier appears to classify some classical short GRBs with extended emission as mergers. However, it is unclear why others like GRB 120308B would be confidently classified as mergers considering that their lightcurves appear more similar to typical long bursts. This demonstrates that our model could be a new avenue for identifying possible candidates for future progenitor confusion studies.

We also reiterate our recommendation that supernovae be searched for after "short" bursts as well as "long", as we suspect from our classifier that there exist some collapsar GRBs whose $\text{T}_\text{90}$ is well below 4.2 seconds. This includes possibly using a similar classifier to predict which progenitor we may detect, rather than sorting bursts as long or short as many instruments currently do. Whether these short collapsars have jets weaker than typical Type Ic supernovae \citep{200826A_weak_jet} or are only the precursor of their events \citep{200826A_geometry}, we know that at least one exists \citep[GRB 200826A][]{short_collapsar_paper}. If that remains the only one found after a long period of searching, it becomes more likely that only short-lived central engines form "short collapsars." If "short collapsars" become more common as they are searched for, it becomes more likely that they are a "common rarity" that requires models to explain the link between the progenitor and the central engine's shorter lifespan. However, it will be difficult to judge without a second event with which to compare GRB 200826A.

\section{Conclusions}\label{conclude}
In this paper, we outline the potential for a GRB progenitor classifier based on standard prompt emission properties. We found that the ratio of the peak energy over the prompt fluence could be preliminarily used as a progenitor classifier when combined with the $\text{T}_{\text{90}}$. It appeared to split bursts at a generally longer duration than the accepted value of 4.2s for the GBM catalog, which we suspect to be due to the presence of "long mergers" \citep{Fermi_Ten-Year_Catalog}. We suggest further studies be performed when more progenitors are correlated with data from instruments like GBM that can perform highly specific spectral studies of burst prompt emission. These studies may significantly affect the classifier, including potentially necessitating the use of more complex functions to explore the boundary or inclusion of more variables as progenitor prediction models become more complex. We also recommend that future studies examine how this relationship changes for the very highest fluence bursts when possible. If the predictions of \cite{Eric's_221009A_paper} are correct, there may be up to 3 of these bursts per decade (assuming that the high-fluence cutoff is $\sim10^{-3}$ erg cm$^{-2}$.

\section*{}
This work made use of data supplied by the UK Swift Science Data Centre at the University of Leicester. We gratefully acknowledge Phil Evans and Israel Martinez Castellanos for helpful discussions. We also thank the referees for their helpful commentary.

\bibliography{bibliography}{}
\bibliographystyle{aasjournal}

\appendix
\section{Processed Data}\label{Data_Table}
\subsection{Known Progenitors}
\begin{longtable}{|c|c|c|c|}
    \caption{The following two tables demonstrate the very unbalanced nature of our known progenitor sample. While there are dozens of GRB-associated supernova detections, even within our time constraints, there are very few GRB-associated mergers. Furthermore, our exotic and "short collapsar" burst progenitors are unique. This limits our ability to interpret extensions of our model to unusual bursts. Even only requiring GBM to view a known progenitor eliminates a large number of known kilonovae and supernovae from Swift-BAT and INTEGRAL. \label{progenitor table}}\\
        \hline
        Name & Alt\_Name & Available Data & Source \\ \hline
         \multicolumn{4}{|c|}{Long Collapsars} \\ \hline
        GRB 091127 & GRB 091127976 & GBM & \cite{supernova_paper} \\ \hline
        GRB 101219B & GRB 101219686 & BAT, XRT, GBM, redshift & \cite{supernova_paper} \\ \hline
        GRB 140606B & GRB 140606133 & GBM & \cite{supernova_paper} \\ \hline
        GRB 150210A & GRB 150210935  & GBM & \cite{2023MNRAS.518.6243J} \\ \hline
        GRB 180720B & GRB 180720598 & BAT, XRT, GBM, redshift & \cite{supernova_paper} \\ \hline
        GRB 180728A & GRB 180728728 & BAT, XRT, GBM, redshift & \cite{supernova_paper} \\ \hline
        GRB 190829A & GRB 190829830 & BAT, XRT, GBM, redshift & \cite{supernova_paper} \\ \hline
        GRB 200826A & GRB 200826187 & GBM & \cite{supernova_paper} \\ \hline
        \multicolumn{4}{|c|}{Short Collapsars} \\ \hline
        GRB 200826A & GRB 200826187 & GBM & \cite{short_collapsar_paper} \\ \hline
\end{longtable}

\begin{longtable}{|l|l|l|}
\caption{Each row gives the reference to a paper that described the observational characteristic that we used to identify the object as a merger, if it did not perform it directly. Some mergers-associations in this table are performed by directly observing the infrared spectrum of r-process nucleosynthesis. Those which are only detected as a bump in the infrared where r-process nucleosynthesis might be creating new elements are labelled as weak kilonova fit. Those short mergers which occurred in the outer reaches of their galaxies are labelled as position inferred. Finally, one short GRB also had a very small spectral lag, and is thus suspected to be a merger.
    \label{kilonova_table}}
    \\ \hline
\textbf{Name} & \textbf{Source} & \textbf{Merger Asso. Method} \\ \hline
    GRB 050509B & \cite{Galactic_Mergers_Paper} & position inferred\\ \hline
    GRB 050709A & \cite{Galactic_Mergers_Paper} & position inferred \\ \hline
    GRB 051210A & \cite{Galactic_Mergers_Paper} & position inferred \\ \hline
    GRB 060614  & \cite{GCN_060614} & kilonova detected \\ \hline
    GRB 070714B & \cite{Galactic_Mergers_Paper} & position inferred \\ \hline
    GRB 070809  & \cite{GCN_070809} & kilonova detected \\ \hline
    GRB 071227A & \cite{Galactic_Mergers_Paper} & position inferred \\ \hline
    GRB 080503  & \cite{GCN_080503A} & weak kilonova fit \\ \hline
    GRB 080905A & \cite{Galactic_Mergers_Paper} & position inferred \\ \hline
    GRB 090515A & \cite{Galactic_Mergers_Paper} & position inferred \\ \hline
    GRB 111005A & \cite{GCN_111005A} & kilonova detected \\ \hline
    GRB 120304B & \cite{GCN_120304B} & spectral lag indicated \\ \hline
    GRB 130603B & \cite{GCN_130603B} & kilonova detected \\ \hline
    GRB 150101B & \cite{GCN_150101B} & kilonova detected \\ \hline
    GRB 160624A & \cite{GCN_200522A} & weak kilonova fit \\ \hline
    GRB 160303A & \cite{Galactic_Mergers_Paper} & position inferred \\ \hline
    GRB 160821B & \cite{GCN_160821B} & kilonova detected \\ \hline
    GRB 170817A & \cite{GCN_170817A} & kilonova detected \\ \hline
    GRB 200522A & \cite{GCN_200522A} & weak kilonova fit \\ \hline
    GRB 211211A & \cite{GCN_211211A} & kilonova detected \\ \hline
    GRB 230307A & \cite{GCN_230307A} & kilonova detected \\ \hline
\end{longtable}

\subsection{Progenitors as Sorted by Classifier}\label{Model 2 Progenitors}
\begin{longtable}{|c|c|c|c|c|}
\caption{This table contains a small smaple of the probabilities that each of our GBM bursts is associated with either a merger or collapsar event, as derived by our SVM algorithm. This algorithm was fitted using a radial kernel on $E_{p, prompt} S_{\gamma, prompt}^{-1}$ versus the $\text{T}_{\text{90}}$ of the known progenitors of \textit{Fermi} GRBs 211211549, 200826187, 190829830, 180728728, 180720598, 171010792, 160624477, 150101641, 150210A, 140606133, 130427324, 130215063, 120304248, 101219686, 091127976, 090618353, and 80905499. As one can see, most bursts strongly prefer one progenitor over the other, leading us to assert that the two-progenitor system is most likely correct for most bursts. As we only fitted a small number of bursts and applied that fit to the much larger sample, this model must be interpreted cautiously. We suggest that future groups investigate adding more progenitors, especially in the "short" collapsar category. For more information, please see "Goldstein\_Classification\_sorted.csv." \label{fig: unnecessarily long table}}
\\ \hline
Designation & $\text{T}_\text{90}$ & $E_{p, prompt} S_{\gamma, \text{prompt}}^{-1}$ & $P_{collapsar}$ & $P_{merger}$ \\ \hline
    GRB 101219686 & 51.009 & 2.07E+07 & 3.56E-01 & 6.44E-01 \\ \hline
    GRB 091127976 & 8.701 & 1.07E+06 & 8.43E-01 & 1.57E-01 \\ \hline
    GRB 171010792 & 107.266 & 2.18E+05 & 5.97E-01 & 4.03E-01 \\ \hline
    GRB 180720598 & 48.897 & 2.13E+06 & 7.06E-01 & 2.94E-01 \\ \hline
    GRB 130215063 & 143.746 & 4.06E+06 & 5.30E-01 & 4.70E-01 \\ \hline
    GRB 130427324 & 138.242 & 3.35E+05 & 4.83E-01 & 5.17E-01 \\ \hline
    GRB 140606133 & 22.784 & 7.61E+07 & 1.89E-01 & 8.11E-01 \\ \hline
    GRB 090618353 & 112.386 & 5.55E+05 & 5.42E-01 & 4.58E-01 \\ \hline
    GRB 120304248 & 5.376 & 1.31E+08 & 4.30E-01 & 5.70E-01 \\ \hline
    GRB 150101641 & 0.08 & 5.25E+08 & 8.96E-01 & 1.04E-01 \\ \hline
    GRB 160624477 & 0.384 & 2.94E+09 & 6.49E-01 & 3.51E-01 \\ \hline
    GRB 120403857 & 4.288 & 1.29E+11 & 2.12E-01 & 7.88E-01 \\ \hline
    GRB 120227725 & 17.408 & 5.87E+06 & 6.91E-01 & 3.09E-01 \\ \hline
    GRB 141205018 & 13.056 & 1.40E+08 & 3.86E-01 & 6.14E-01 \\ \hline
    GRB 180630467 & 12.032 & 2.57E+07 & 8.41E-01 & 1.59E-01 \\ \hline
    GRB 170116238 & 9.216 & 1.62E+08 & 9.14E-01 & 8.64E-02 \\ \hline
    GRB 091026550 & 8.96 & 3.48E+08 & 7.44E-01 & 2.56E-01 \\ \hline
    GRB 150312403 & 0.32 & 2.15E+09 & 9.52E-01 & 4.80E-02 \\ \hline
    GRB 110520302 & 12.288 & 2.81E+08 & 6.20E-01 & 3.80E-01 \\ \hline
    GRB 121216419 & 9.216 & 1.52E+09 & 8.91E-01 & 1.09E-01 \\ \hline
    GRB 101227195 & 95.488 & 6.63E+07 & 5.67E-01 & 4.33E-01 \\ \hline
    GRB 160326062 & 19.456 & 1.69E+08 & 5.55E-01 & 4.45E-01 \\ \hline
    GRB 101116481 & 0.576 & 1.44E+09 & 2.22E-01 & 7.78E-01 \\ \hline
    GRB 170130697 & 29.184 & 2.34E+08 & 8.26E-01 & 1.74E-01 \\ \hline
    GRB 110205027 & 5.376 & 8.44E+08 & 7.22E-01 & 2.78E-01 \\ \hline
\end{longtable}

\section{Error Propagation} \label{Error Bars}
The model-dependent uncertainty in the peak energy is given in the GBM catalog. We assumed that the spectroscopic model was correct, though this is itself known to have a level of uncertainty. The prompt fluence uncertainty was also taken from the catalog, as above, making the error bars in the peak energy over the fluence:
\begin{equation}\sigma_{\frac{E_P}{S}}=\sqrt{\left(\frac{\sigma_{E_P}}{S}\right)^{2}+\left(\frac{E_P\sigma_{S}}{S^{2}}\right)^{2}}=\left(\frac{E_P}{S}\right)\sqrt{\left(\frac{\sigma_{E_P}}{E_p}\right)^{2}+\left(\frac{\sigma_{S}}{S}\right)^{2}}\end{equation}
where each error is taken to be the average of its left and right values. The error in $\text{T}_{\text{90}}$ could be taken directly from the GBM catalog. 
\begin{figure}[h!]
    \centering
    \includegraphics[width=0.75\textwidth]{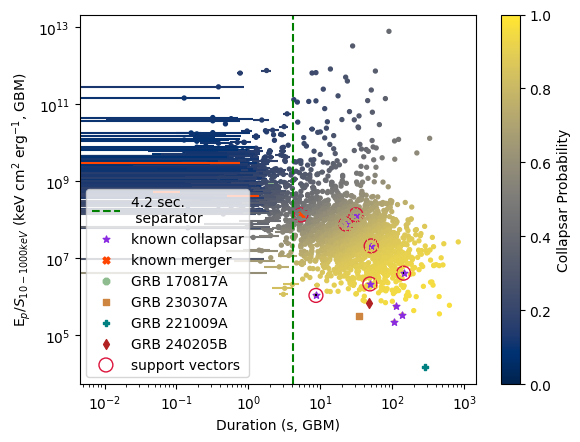}
    \caption{In this version of the sorted data, it is all but impossible to view the progenitor classes concerning the geometric division of the data. It also becomes much more obvious that the uncertainty in this model is very high. This model only worked as well as it did because GBM (and \textit{BATSE} before it) can specify the peak energy of the prompt much better than BAT. We caution that this model should not be applied using instruments with narrow prompt bandpasses that cannot analyze the prompt spectrum in as much detail. In our testing, it did not work well, if at all, with even a relatively large sample from BAT of bursts that were detected as having a cutoff power law.}
    \label{fig:Goldstein_error_bars_plot}
\end{figure}
For the figure that records the prompt fluence versus the afterglow flux at 11 hours, we again took the prompt fluence to have the GBM cataloged uncertainty. We estimated the error in the X-ray flux at 11 hours as the change in the average flux. This should have been a clear overestimation, but even with three points recorded as having fluxes well below what XRT can nominally measure, their uncertainties were not noticeably different from their neighbors. 
\end{document}